\documentclass[prb,superscriptaddress,twocolumn,nofootinbib]{revtex4-2}
\usepackage[utf8]{inputenc}
\usepackage{booktabs}
\usepackage{makecell}
\usepackage{verbatim}
\usepackage{graphicx}
\usepackage{xcolor}
\usepackage{amsmath}
\usepackage{amsfonts}
\usepackage{amsthm}
\usepackage{amssymb}
\usepackage{mhchem}
\usepackage{pifont}
\usepackage{amsbsy}
\usepackage{wasysym}
\usepackage{bm}
\usepackage{mathrsfs}
\usepackage{color}
\usepackage{hyperref}
\usepackage{soul}
\usepackage{multirow}
\usepackage{bbm}
\usepackage{physics}
\usepackage{booktabs}
\usepackage[margin=1in]{geometry}
\usepackage{dutchcal}
\usepackage{float}
\usepackage{appendix}
\usepackage[percent]{overpic} 
\usepackage{subcaption}
\hypersetup{
    colorlinks=true,
    linkcolor=red,        
    citecolor=blue,
    breaklinks=true,
    urlcolor=blue
}
\let\saveboldmath\boldmath
\usepackage{mathptmx}
\let\boldmath\saveboldmath

\DeclareMathAlphabet{\pazocal}{OMS}{zplm}{m}{n}

\DeclareSymbolFont{cmsymbols}{OMS}{cmsy}{m}{n}
\SetSymbolFont{cmsymbols}{bold}{OMS}{cmsy}{b}{n}
\DeclareSymbolFontAlphabet{\mathcalB}{cmsymbols}
\newcommand{\pazocalB}[1]{\bm{\mathcalB{#1}}}

\begin{document}

\preprint{APS/123-QED}

\title{Static and Dynamical Characterization of Ground State Phases Induced by Frustration and Magnetic Field in the Spin-1 Orthogonal Dimer Chain }

\author{Ernest Ong}
\thanks{These authors contributed equally to the work}
 \affiliation{School of Physical and Mathematical Sciences, Nanyang Technological University, Singapore}
\author{Dhiman Bhowmick}
\thanks{These authors contributed equally to the work}%
 \affiliation{Department of Physics, National University of Singapore, Singapore}
\author{Sharoz Schezwen}
\thanks{These authors contributed equally to the work}
\affiliation{School of Physical and Mathematical Sciences, Nanyang Technological University, Singapore}%
\author{Pinaki Sengupta}
\affiliation{School of Physical and Mathematical Sciences, Nanyang Technological University, Singapore}%

\date{\today}

\begin{abstract}

The spin-$1$ orthogonal dimer chain is investigated using the Density Matrix Renormalization Group (DMRG) algorithm. A transformation to a basis that uses the local eigenstates of the orthogonal dimers, while retaining the local spin states for the parallel spins, allows for more effective implementation of the symmetries, as well as mitigating the entanglement bias of DMRG. A rich ground state phase diagram is obtained in the parameter space spanned by the ratio of inter- to intra-dimer interaction (which measures the degree of frustration) and an external magnetic field. Some ground state phases exhibit effective Haldane chain character, whereas others exhibit fragmentation of the ground state wavefunction, or clustering. The phases are characterized by their static properties, including (local) spin quantum number, entanglement entropy, and the spin-spin correlation function. Detailed characterization of a carefully selected set of representative states is presented. The static properties are complemented by exploring the low-energy dynamics through the calculation of the dynamic structure factor. The results provide crucial insight into the emergence of complex ground state phases from the interplay between strong interactions, geometric frustration, and external magnetic field for interacting S=1 Heisenberg spins.

\end{abstract}

\maketitle


\section{\label{sec:level1}Introduction}

Geometrically frustrated quantum spin chains provide a versatile platform to study quantum many-body physics.  
The interplay between competing interactions, geometric frustration, and enhanced quantum fluctuations due to lower dimensionality results in a wide range of unconventional ground state phases and anomalous magnetization that are not observed in their non-frustrated and higher-dimensional counterparts. 
Availability of powerful analytical and numerical methods in one dimension allows us to study the 
emergence of these novel phases and probe their properties in a well-controlled manner. 
Two prominent and extensively studied examples of frustrated spin chains include the spin-diamond chain and the 
orthogonal dimer chains. 
There exist several quasi one-dimensional (1D) quantum magnets where many of 
such novel phases can be realized and studied controllably. The ability to tune many of these 
phases using an external magnetic field makes it possible to characterize them thoroughly. 
The study of these systems is also important for practical reasons -- the unique functionalities 
associated with many of these states can potentially be harnessed for next generation
of technological breakthroughs. This makes the understanding, and more importantly, controlling the 
emergence of novel quantum phases in quantum spin chains an active frontier of contemporary condensed
matter physics.

A unique feature that is observed in many frustrated spin chains is the fragmentation of the ground
state into clusters of spins. Within each cluster, the spins form a unique arrangement of singlets and, 
for states with net magnetization, bound spin states with finite moment (e.g., triplons). The ground
state is a direct product of these clusters. Experimental signatures of fragmentation in several 
quantum magnets have provided further stimulus, driving the field. A direct consequence of such 
ground state fragmentation is the appearance of magnetization plateaus in the presence of an external 
magnetic field - a unique feature observed in many low-dimensional frustrated spin systems. A formal 
condition for the emergence of magnetization plateaus in 1D Heisenberg model 
was derived in Ref.\onlinecite{Oshikawa1984}, which predicts that the number of plateaus increases with the size of the clusters. It was shown independently that an infinite sequence of plateaus is stabilized for the spin-${1\over 2}$ orthogonal dimer chain. The theoretical studies are backed up by experimental observation of magnetization plateaus in frustrated spin chain compounds, such as azurite.

Such spin clusters were also found in specially tuned spin-$\frac{1}{2}$ ladder systems, for both even and odd-numbered chain lengths \cite{sugimoto, pollman}. Although the length of the chain studied plays a role in the creation of different phases, conditions can be made to give an easier comparison. Specifically, the even degeneracy from Kramer's theorem can be lifted by enforcing conditions that allow the odd-numbered systems to mimic the creation of clusters similar to even-numbered systems in systems with the Heisenberg Hamiltonian \cite{sugimoto}. 

While most studies on frustrated spin chains have focused on spin-${1\over 2}$ systems, spin-$1$ systems offer an even richer array of phases due to their expanded Hilbert space. In his seminal
work, Haldane established that the ground state of an antiferromagnetic spin-1 (any integer spin) Heisenberg chain has a finite gap to the lowest magnetic excitation, in contrast to their spin-${1\over 2}$
(any half-odd integer) counterparts. The nature of the ground state for the spin-1 chain was 
elucidated by Affleck, Kennedy, Tasaki, and Lieb (AKLT).  Through exact solution of the bilinear-biquadratic model at a special point, they showed that the ground state of the spin-1 Heisenberg chain, commonly referred to as the Haldane phase, is adiabatically connected to a valence bond solid comprised of, on every nearest neighbor bond, a singlet formed from two pseudo-spin-${1\over 2}$ moments arising from the symmetric (mathematical) decomposition of spin-1 moments at each site. Thus, ground state fragmentation occurs even in the canonical Heisenberg chain for spin-1 moments. The particular structure of the Haldane phase results in symmetry-protected gapless edge states, making it one of the earliest and most extensively 
studied examples of symmetry-protected topological (SPT) states. Experiments on quasi-1D spin-1 quantum magnets such as NENP, \ce{Y2NiO5}, and \ce{AgVP2S6} have confirmed the theoretical predictions. 

Relatively few works exist on frustrated spin-1 chains. The spin-1 (and other spin $\ge {1\over 2}$) orthogonal dimer chain was studied in Ref.~\onlinecite{src0b032}, where a series of gapped phases separated by first-order transitions were uncovered as the ratio between intra- and inter-dimer interaction is varied. Transformation to a basis of local eigenstates of the orthogonal dimer
was introduced in ref.~\onlinecite{Takano2017} for the spin-1 diamond chain. More recently, the mixed spin-$1$ spin-$\frac{1}{2}$ Heisenberg octahedral chain has shown a remarkably rich phase diagram consisting of gapless, gapped, and spin liquid states, in the presence of an external magnetic field. Tuning the field showed the existence of multiple Haldane-type clustered phases with different periodicities and static properties \cite{Katarina}.

In this paper, the one-dimensional spin-$1$ orthogonal dimer chain (Figure \ref{fig:orthogonalDimerStructure}) is studied in odd chain lengths using the Density Matrix Renormalization Group (DMRG) method, where a rich phase diagram consisting of multiple Effective Haldane Phases, Haldane-type clustered phases, and phases with low entanglement and localized excitations was found.

This paper is ordered as follows
\begin{itemize}  
    \item Section 2 details the Hamiltonian construction, the basis transformation, and the methods to obtain and extrapolate the Dynamical Structure Factor Plots
    \item Section 3 details the creation of the phase diagram, and explores the different phase types found
    \item Section 4 provides a summary and concluding remarks
\end{itemize}

\section{\label{sec:level1}Model and Methodology}\label{2}

\subsection{System Hamiltonian}\label{dbasis}

\begin{figure}
    \centering
    \includegraphics[width=1\linewidth]{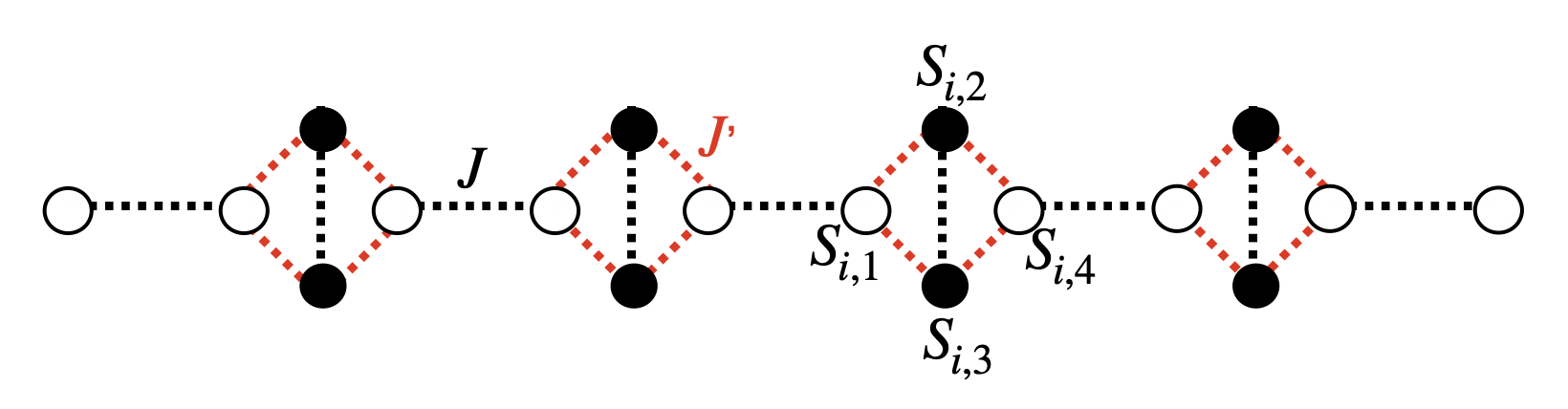}
    \caption{The spin-$1$ orthogonal dimer chain. The filled (open) sites form the perpendicular dimers (parallel sites). The intra- and inter-dimer interactions are denoted by $J$ and $J'$ respectively.} 
    \label{fig:orthogonalDimerStructure}
\end{figure}

We study a spin-$1$ orthogonal dimer chain (Fig.\,\ref{fig:orthogonalDimerStructure}), which consists of parallel sites (open circles) and orthogonal dimers (filled circles). The inter-dimer interaction induces frustration in the chain. The Hamiltonian describing the system is given by,
\begin{equation}
\begin{split}
    &\hat{H}  = J \sum_{i=1}^N \left[\hat{S}_{i,2}\cdot\hat{S}_{i,3}+\hat{S}_{i,4}\cdot\hat{S}_{i+1,1}\right] -h\sum_{i=1}^N \sum_{\alpha=1}^{4}\hat{S}_{i,\alpha}^z
    \\
    &+J' \sum_{i=1}^N \sum_{\alpha\in(1,3)} \hat{S}_{i,\alpha}\cdot\hat{S}_{i,\alpha+1}
    +
    J' \sum_{i=1}^N \sum_{\alpha\in(1,2)} \hat{S}_{i\alpha}\cdot\hat{S}_{i,\alpha+2},
    \label{eqn:Hamiltonian}
\end{split}
\end{equation}
where $J$ and $J'$ are the intra- and inter-dimer interaction strengths.
The indices $i$ and $\alpha$ represent the unit cell and sub-lattice, respectively. 
 
To perform a more effective simulation of the model, a mapping was formulated for the orthogonal dimers, to transform the quasi-$1$D chain into effectively a true one dimensional chain (the construction of the new basis for the dimers is explained in Section \ref{dbasis}). By performing this transformation, the system encodes the symmetries of the Hamiltonian in a better way
, by reducing the bias caused by mapping a higher dimension model to a quasi-$1$D chain typically done in the DMRG algorithm. The origin of this bias stems from the short-range interactions being transformed into long-range interactions in the DMRG algorithm, producing a convergence to a ground state that minimizes the entanglement specific to the chosen $1D$ configuration. 
To mitigate this, we apply an exact basis transformation, which allows us to map the system onto a $1D$ chain without altering the short-range nature of the interactions, which can be done via a basis transformation on the dimer orthogonal to the chain, mediated by the unitary matrix $U$,
\begin{equation}
    \ket{s_t, m_t}=\sum_{m_1, m_2} U_{\lbrace s_t, m_t\rbrace, \lbrace m_1, m_2\rbrace}\ket{m_1, m_2},
\end{equation}
where $m_i$ denotes the $\hat{S}^z$ quantum number on the $i$-th spin within the dimer, while ($s_t$,$m_t$) denote the total spin $\hat{S}$ and the total $\hat{S}^z$ quantum number of the orthogonal dimer respectively.
The elements of the unitary matrix $U$ are the Clebsch–Gordan coefficients.
Under the basis transformation,  the single spin operators of the orthogonal dimers transform as,
\begin{equation}
    \hat{S}^{\alpha}_i\rightarrow\sum_{t, t'}
    \bra{s_t, m_t}\hat{S}^{\alpha}_i\ket{s_{t'}, m_{t'}}
    \hat{t}^\dagger \hat{t}',
\end{equation}
where, $\hat{t}^\dagger$ (or $\hat{t}$) is the creation (or annihilation) operator at the state $\ket{s_t, m_t}$, and $\hat{S}_i^\alpha$ denotes the $\alpha$-component of spin operator at the $i$-th site of the dimer.
In the literature, such a transformation is called as bond-operator transformation\,\cite{BondOperator1, BondOperator2, BondOperator3, BondOperator4}.

Upon basis transformation, a uniform magnetic field is applied in the form of a Zeeman term (Eq. \ref{eqn:Hamiltonian}), where the system evolves to exhibit a wide variety of ground state phases. To visualize the periodicities of the phases, the magnetization and site quantum number (SQN) were calculated for each site on parallel bonds (hollow circle in Figure \ref{fig:orthogonalDimerStructure}, henceforth referred to as parallel sites) and for each orthogonal dimer. 

To elucidate the nature and extent of the topological properties of the phases found, we have also probed the values of the nonlocal string order parameter in this paper \cite{stringorder}. In the small perturbation regime, the Kramer's degeneracy is lifted when a small magnetic field is applied that breaks the Time-Reversal Symmetry. As a consequence, a finite string order has been shown to be related to the existence of a topological phase transition. Therefore, for the phases investigated, we performed the calculation of the string order parameter in Equation \ref{so} using periodic boundary conditions, to characterize the types of phases found (results given in Figure \ref{fig:de-en-ee}$c$).

\begin{equation}
    \label{so}
\mathcal{O}_{\text{string}}^{z} = \lim_{|i-j| \to \infty} \left\langle S_i^z \exp\left( i \pi \sum_{k=i+1}^{j-1} S_k^z \right) S_j^z \right\rangle
\end{equation}


\captionsetup{justification=justified, font=small}

\subsection{Dynamical Structure Factor (DSF)}\label{dsfsection}

For further insight into the nature of the different ground state phases, we have probed their low energy excitation spectrum by calculating the Dynamic Structure Factor (DSF). The DSF is defined by 
\begin{equation}
\begin{split}
    \pazocal{S}^{\alpha\beta}&(k,\omega) = \sum\limits_{n=-\infty}^{\infty} e^{-ikn} \int_{-\infty}^{\infty}
    dt\;e^{i\omega t} \; \pazocal{O}^{\alpha\beta}_n(t)
    \\
    &\text{where, } \pazocal{O}^{\alpha\beta}_n(t)=\langle\psi|\;\hat{S}^{\alpha}_n(t)\hat{S}_0^{\beta}(0)\;|\psi\rangle
\end{split}
    \label{eqn:DSF}
\end{equation}
where, we consider $\alpha=\beta=z$ unless otherwise stated. This provides information on the low-energy longitudinal excitation.
The autocorrelation function $\pazocal{O}^{\alpha\beta}_n(t)$ is evaluated using time-evolving block decimation (TEBD) with a second order Trotter decomposition.
For the evaluation of the DSF, one requires autocorrelation functions over a long range of times to approximate the improper integral with time interval $t\in(-\infty, +\infty)$.
Nevertheless, TEBD is limited in performing time evolution over extended durations, as the associated error in the time-evolved state—$O(T\delta\tau^2)$ for a second-order Trotter decomposition—grows proportionally with both the square of the time step, $\delta\tau^2$, and the total evolution time, $T$.
Calculating the DSF using data collected over a short time range using TEBD results in severe ringing artifacts \,\cite{bartelLinPred}, leading to a low-resolution output with no discernible features.

Therefore, we apply {\it linear prediction} along with TEBD to improve the DSF calculation\,\cite{bartelLinPred}.
The {\it linear prediction} method is a pedagogical machine learning technique which we use to extrapolate the autocorrelation function $\pazocal{O}_n^{\alpha\beta}(t)$—obtained from TEBD—beyond its original time range.

In this paper, the ground state calculation before the TEBD simulation was calculated with a bond dimension of 800, and for 200 sweeps, for a total time of $t=10$. The TEBD simulation was done using this ground state with a timestep of $\tau = 0.001$. We then employed the {\it linear prediction} method to extrapolate the signal until 20,000 timesteps.

Given a TEBD dataset in time $\pazocal{O}(t_1), \pazocal{O}(t_1), \ldots, \pazocal{O}(t_{N_t}) $, that is equidistant in the time axis, we can predict the data at future times $\pazocal{O}(t_{{N_t}+1}), \pazocal{O}(t_{{N_t}+2}), \ldots$, using the {\it linear prediction} ansatz,
\begin{equation}
    \pazocal{O}(t_{n+1})=-\sum_{i=0}^{p-1} a_i \pazocal{O}(t_{n-i}),
\end{equation}
where $p$ are the previous values in the time series and $a_i$ are the coefficients determined using known data.
We can represent this equation in a matrix form,
\begin{equation}
\pazocalB{O}_{n+1}=A\pazocalB{O}_n     
\end{equation}
where the matrix $A$ is a $p\times p$ matrix given by,
\begin{equation}
    A =
    \begin{bmatrix}
    -a_0 & -a_1 & -a_2 & \dots & -a_{p-1}\\
    1 & 0 & 0 & \dots & 0 \\
    0 & 1 & 0 & ... & 0 \\
    \vdots & \ddots & \ddots &\ddots & \vdots \\
    0 & \dots & 0 &1 & 0
    \end{bmatrix}.
    \label{eqn:A}
\end{equation}
Unlike generic machine learning techniques, the coefficients $a_i$ in {\it linear prediction} can be determined analytically using the following expression,
\begin{equation}
    \boldsymbol{a} = -\boldsymbol{R}^{-1}\boldsymbol{r},
\end{equation}
where $\boldsymbol{R}$ and $\boldsymbol{r}$ are the autocorrelation matrices as,
\begin{equation}
    R_{ji} = \sum_{n} \frac{\pazocal{O}^*(t_{n-j})\pazocal{O}(t_{n-i})}{\omega_n},
\;\;\;\;
    r_{j} = \sum_{n} \frac{\pazocal{O}^*(t_{n-j})\pazocal{O}(t_{n})}{\omega_n}.
\end{equation}
Here, $n$ denotes data points within the TEBD data set used to compute the autocorrelation matrices.
Whereas, the indices $i$ and $j$ ranges from $0$ to $p-1$.
Moreover, we choose the weight $\omega_n$ to be $1$.

Recursively applying this matrix $m$-times on the TEBD data of $N_t$-time slices, we can generate $m$ extrapolated data as,
\begin{equation}
\pazocalB{O}_{N_t+m}^{\text{prediction}}=A^m\pazocalB{O}_{N_t}     
\end{equation}
This problem can be further simplified into a problem of diagonalization of the matrix $A$ as,
\begin{equation}
    \pazocal{O}(N_t+m)
    =
    \sum_{i=0}^{p-1} c_i \alpha_i^m,
    \label{eqn:prediction}
\end{equation}
where, $\alpha_i$ are the eigenvalues of matrix $A$ and $c_i=\sum_j (P)_{N_ti} (P^{-1})_{ij} (\pazocalB{O}_{N_t})_j$.
Here matrix $P$ diagonalizes $A$ as $A=P\boldsymbol{\alpha}P^{-1}$ with $\boldsymbol{\alpha}$ being the diagonal matrix with elements as eigenvalues.

However, when $|\alpha_i| > 1$, {\it linear prediction} leads to an exponential growth of the autocorrelation function over time (see Eq.\,\ref{eqn:prediction}).
But, in typical quantum systems, the autocorrelation function is expected to exhibit damped oscillations over time.
There are several ways to address this issue by re-normalizing the spurious eigenvalue to bring it back within the unit circle in the complex plane,
\begin{equation}
    |\alpha_i|>1
    \rightarrow
    |\alpha_i|\leq 1.
\end{equation}
We tested all the following three options ; setting
$\alpha_i\rightarrow{\alpha_i/|\alpha_i|}$,
$\alpha_i\rightarrow 1/\alpha_i^*$, and 
$\alpha_i\rightarrow 0$ for any spurious eigenvalues of $\alpha_i>1$.
 Ultimately, we selected the last one, as it yielded more reliable predictions\,\cite{bartelLinPred}.

\section{\label{sec:level1}Numerical Results and discussion}\label{results}

We have used the Density Matrix Renormalization Group (DMRG) algorithm to simulate the spin-$1$ orthogonal dimer model in finite chains. The DMRG algorithm is implemented using the open-source library ITensor \cite{itensor, itensor2}.  We consider chains with odd number of dimers, with the edges forming parallel sites, as shown in Fig.\,\ref{fig:orthogonalDimerStructure}.
This configuration maintains the inversion symmetry of the system. Results are presented for chains with 155 sites ( $mod \;3 \;=\; 2$ ), with a bond dimension of 500, and 200 DMRG sweeps, to achieve errors smaller than $10^{-8}$.
The ground state phases obtained in the different parameter regimes are characterized according to their static properties. By varying the magnetic field and tuning the interaction strength $J'$, these phases were subsequently separated into three categories in Section \ref{lep},\ref{ehp} and \ref{chp}. It is worth highlighting that Ref.,\cite{Katarina} reports magnetic and frustration-induced clustered Haldane phases in a complex mixed-spin octahedral chain, whereas our findings reveal comparable phases emerging in the present model which 
has attracted considerable interest in the past as it represents the 1D limit of the widely studied Shastry-Sutherland model.
Moreover, in contrast to Ref.,\cite{src0b032}, the present study extends the analysis to include a ferromagnetic frustration parameter $J'$, as well as finite magnetic fields. 
This extended framework uncovers additional exotic phases, such as the clustered Haldane phases, which emerges at experimentally accessible magnetic field strengths.

\captionsetup{justification=justified, font=small}
\begin{figure*}
    \centering
    \begin{overpic}[width=1.0\textwidth,grid=false]{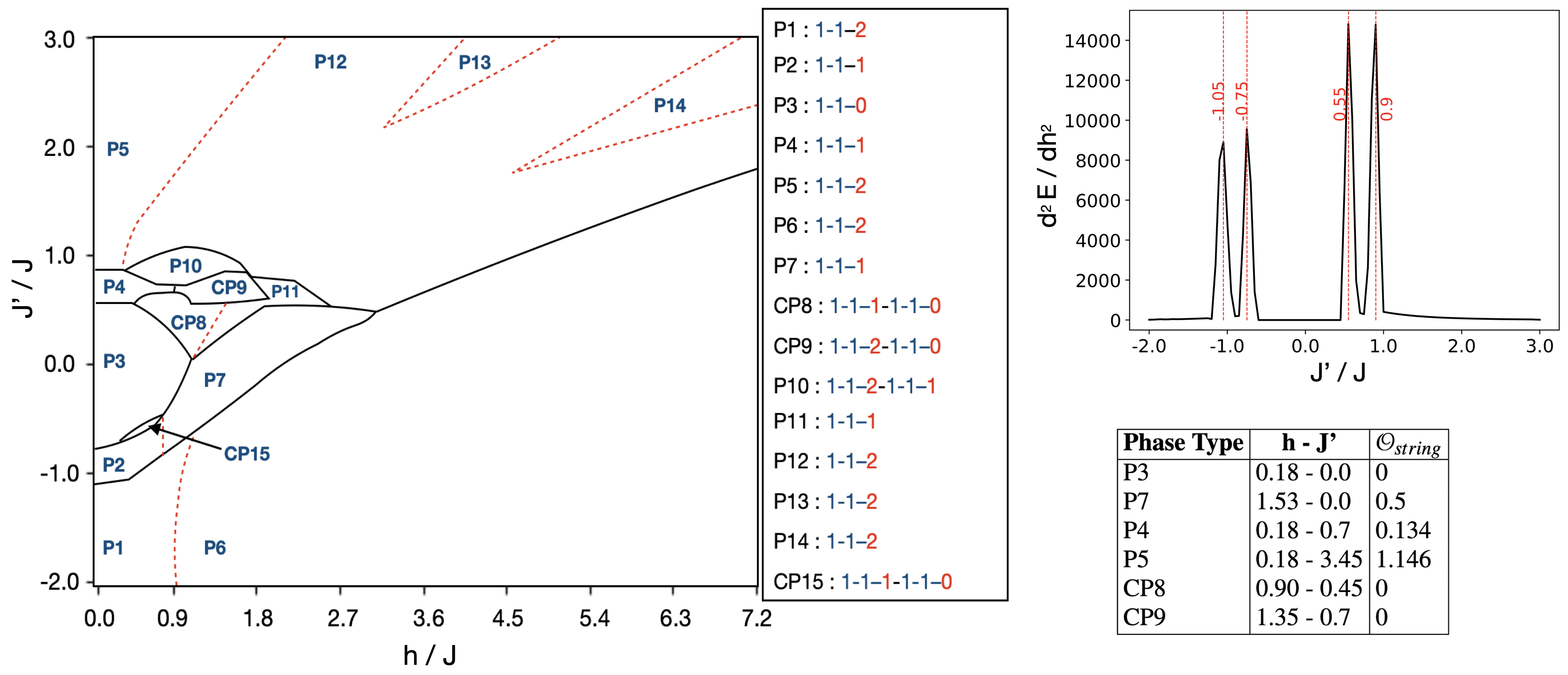} 
    \put(-1,42){\textcolor{red}{\textbf{(a)}}} 

    \put(67,43){\textcolor{red}{\textbf{(b)}}}
    \put(67,15){\textcolor{red}{\textbf{(c)}}}

  \end{overpic}
    \caption{a) The ground phase diagram in the parameter space spanned by the ratio of the inter- to intra-dimer interaction strength ($J'/J$) and an external magnetic field ($h/J$). The red dashed (solid black) lines show a first order (second order) phase transition. The phases are classified  by their total spin quantum number (SQN) as shown in the box to the right where blue (red) digits represent parallel bonded sites (orthogonal dimers). Prefix $CP$ denotes fragmented phases with unit clusters separated by the singlet orthogonal dimers $S=0$ while prefix $P$ refer to non-fragmented phases. 
    b) The second order energy derivative as a function of the interaction strength $\frac{J'}{J}$ at magnetic field $\frac{h}{J}=0.18$. The peaks of the second order derivative indicate quantum phase transitions. c) String Order parameter values for representative phases. The parameter sets represent phases discussed in detail in Sections \ref{lep},\ref{ehp},\ref{chp}. The Effective Haldane phases, $P4$ and $P5$ have finite string order as expected. The phase $P7$, despite having a finite string order, is not consistent with a Haldane-type phase. }
    \label{fig:de-en-ee}
\end{figure*}

\subsection{Phase Diagram Construction}\label{pdconstruct}

A representative phase diagram constructed for the Spin-$1$ orthogonal dimer model is shown in Figure \ref{fig:de-en-ee}$a$. The red dashed (solid black) lines indicate a first-order (second-order) quantum phase transition. 
The different phases found are numbered $1 - 15$, and the corresponding spin quantum number (SQN) is indicated in the box next to fig.~\ref{fig:de-en-ee}$a$. 
The SQN values of the sites on parallel (orthogonal dimer) bonds are shown in blue (red).
In the transformed basis, the parallel sites carry spin-$1$ sites, while the orthogonal dimers can have spin-$=0,1,2$.
The phase diagram is constructed via enumerating first and second order derivatives of energy across a range of magnetic field $\frac{h}{J} = [0.0,7.2]$ and frustration strengths $\frac{J'}{J} = [-2.0,3.0]$.


\subsection{Low Magnetic Field}

We begin by discussing a section of the phase diagram at a small magnetic field, specifically for $\frac{h}{J} = 0.18$, while varying only the frustration parameter $J'/J$.
The phase diagram of this model in the absence of a magnetic field has previously been studied using analytical methods and exact diagonalization, as reported in Ref. \cite{src0b032}. Our results from DMRG calculations are in good agreement with these earlier findings.
Additionally, we also extend our study to negative values of frustration parameter $J'$ and observed five kinds of phases at low magnetic field. 
The phases $P3$, $P4$ and $P5$ are the singlet-dimer, Haldane and Plaquette ordered phases respectively, which are also described in Ref. \cite{src0b032}.
The phases $P1$ and $P2$ found in our study for ferromagnetic values of $J'$, are particularly intriguing, as they are distinct from the fully saturated phase $P6$, despite exhibiting ferromagnetic characteristics.
The emergence of three distinct phases, rather than a single fully polarized phase, can be attributed to the competition between the antiferromagnetic interaction $J$ and the ferromagnetism-favoring interactions $J'$ and $h$.
The transition from the $P1$ to $P2$ is accompanied by a change in SQN of the orthogonal dimer from $2$ to $1$, whereas the transition from $P1$ to $P6$ is accompanied by change in other static properties of the chain, including the magnetization (see Appendix \ref{diff}).






We choose some representative phases in this extensive phase diagram for detailed characterization. These can be broadly characterized as the Low Entanglement Phases, the Effective Haldane Phases and the Clustered Haldane Phases.
The next three subsections explores each category and provide representative static properties, as well as the low energy excitation spectrum.

\subsection{Low Entanglement Phases}\label{lep}

\captionsetup{justification=justified, font=small}
\begin{figure*}
  \centering
  \begin{overpic}[width=0.9\textwidth,grid=false]{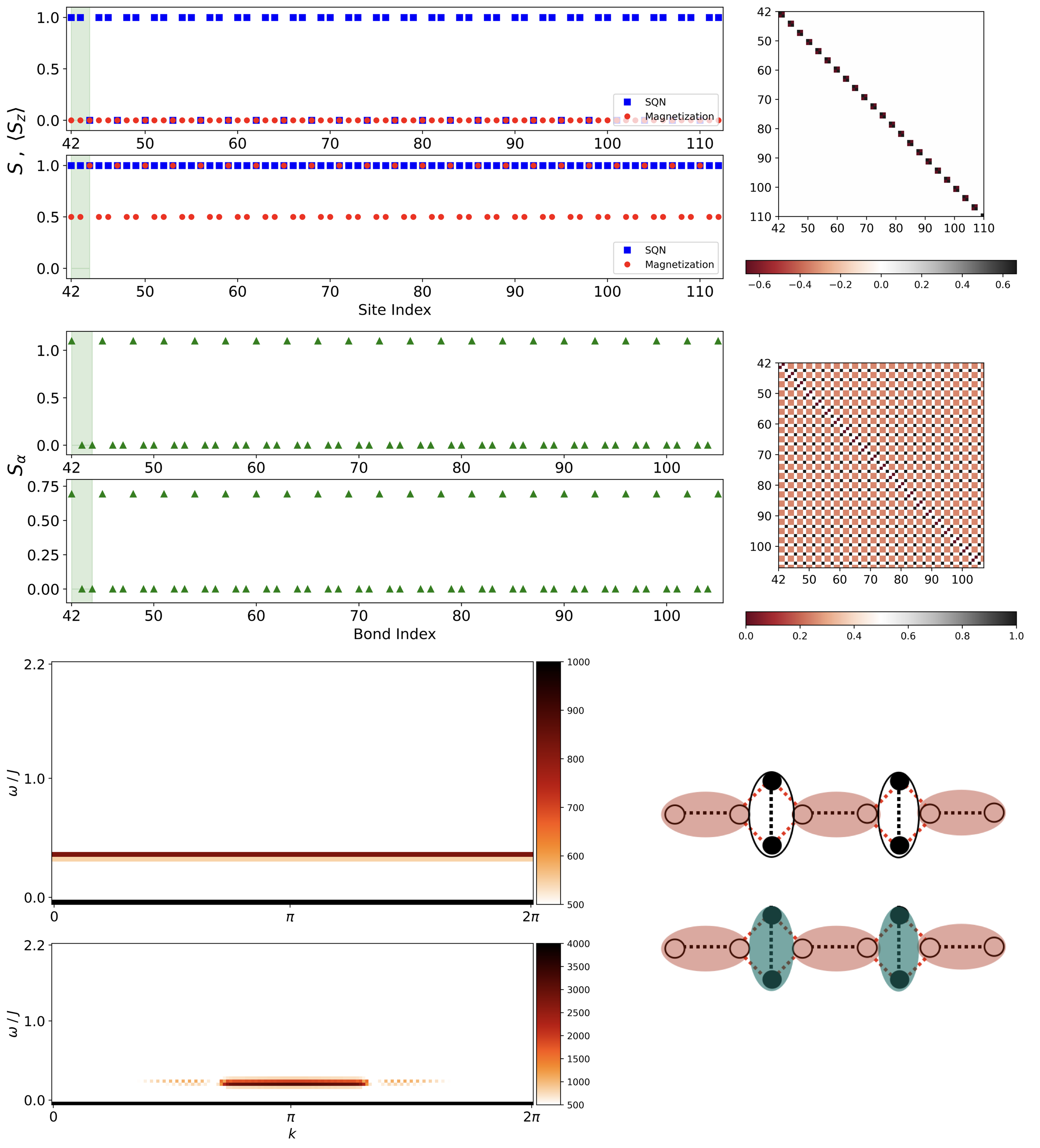} 
    \put(-2,98.5){\textcolor{red}{\textbf{(a)}}} 
    \put(-2,84.5){\textcolor{red}{\textbf{(b)}}}
    \put(-2,70){\textcolor{red}{\textbf{(c)}}}
    \put(-2,57){\textcolor{red}{\textbf{(d)}}}

    \put(63.5,98.5){\textcolor{red}{\textbf{(e)}}}
    \put(63.5,68){\textcolor{red}{\textbf{(f)}}}

    \put(-2,42){\textcolor{red}{\textbf{(g)}}}
    \put(-2,17.5){\textcolor{red}{\textbf{(h)}}}

    \put(60,32.5){\textcolor{red}{\textbf{(i)}}}
    \put(60,20.5){\textcolor{red}{\textbf{(j)}}}
  \end{overpic}
  \caption{a),b) The SQN and magnetization calculations for the phase $P3$ and $P7$ as a function of the site index respectively. c,d) The value of the entanglement entropy as a function of the bond index, for the phase $P3$ and $P7$ respectively. The green boxes indicate the unit cell, beginning with a parallel site. e),f) The correlation matrix for the phase $P3$ and $P7$ respectively. g),h) The dynamical structure factor plot for the phases  $P3$ and $P7$.
  i,j) The representative diagram for the dimers in the phases $P3$ and $P7$ respectively, where the hollow ovals denote singlets, the green ovals denote spin-$1$ triplets, and the pink ovals denote the parallel sites.}
  \label{fig:le}
\end{figure*}
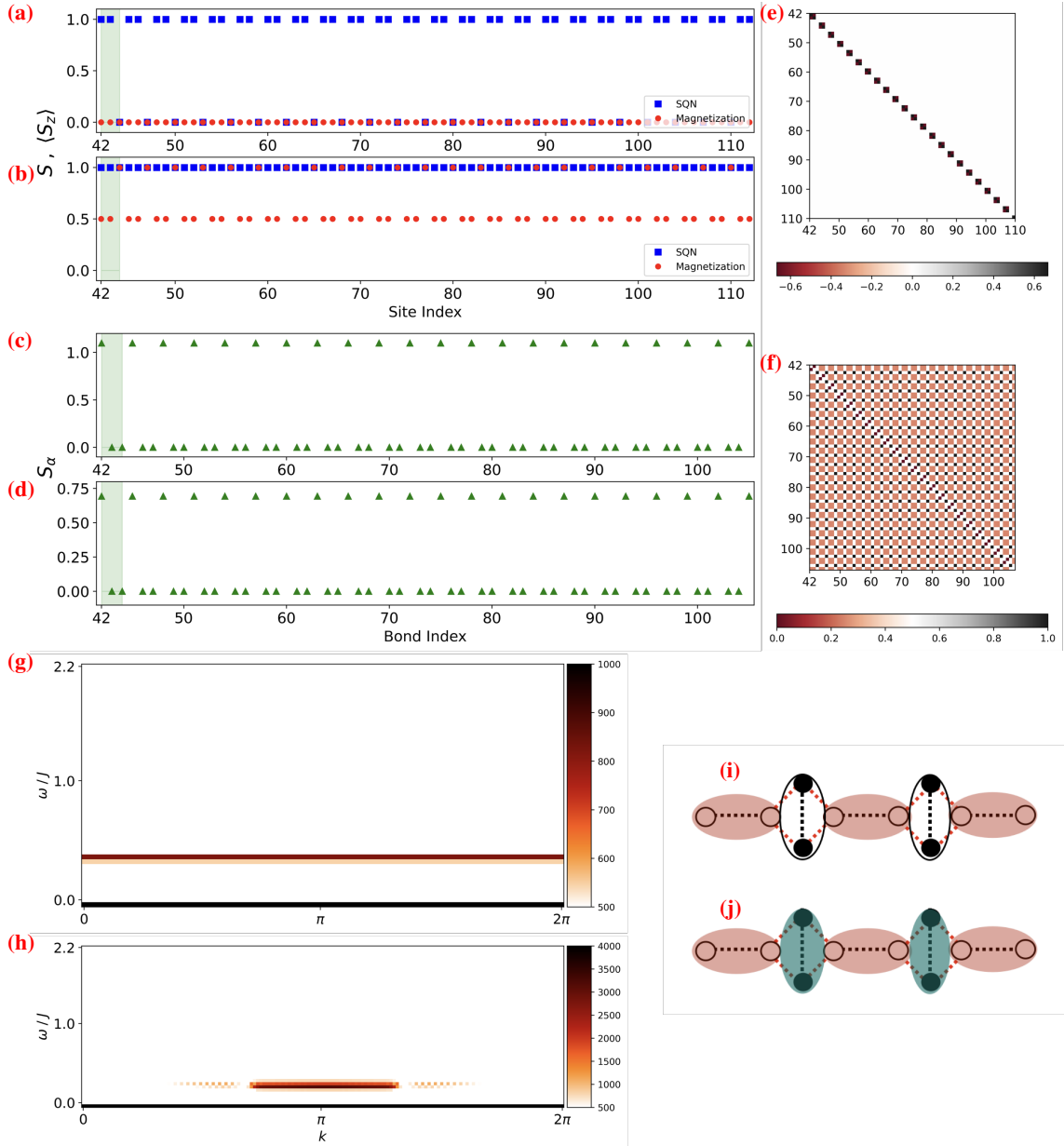

In this section, we discuss the low-entangled phases, $P3$ and $P7$ in Fig.\,\ref{fig:de-en-ee}(a), which can trivially be represented as product states. 
The static and dynamical features of these phases in the bulk are shown in Figures \ref{fig:le}(a)–(f).
In this case, the features are captured between the $42^{\text{nd}}$ and $112^{\text{th}}$ sites of a 155-site system to exclude edge effects.
The green shaded region denotes a unit cell of the system, and the horizontal axis begins with the index of the site in the bulk.
The starting point of the plots, the $42^{\text{nd}}$ site, is the left site of a parallel bond in Fig. \ref{fig:orthogonalDimerStructure}.

The phase $P3$ is a product state of singlets $\frac{1}{\sqrt{3}}\left(\ket{\uparrow\downarrow}+\ket{\downarrow\uparrow}-\ket{00}\right)$ on each orthogonal dimer and the intervening parallel sites (schematic given in Fig. \ref{fig:le}(i)).
This is reflected in the magnetization calculations in Fig.\,\ref{fig:le}(a), where a
zero magnetization is observed throughout the bulk. 
The SQN of each orthogonal dimer is calculated to be zero.
Moreover, there is no entanglement between the orthogonal dimer bonds as seen in Fig.\,\ref{fig:le}(c) (two repeating green triangles at $S_{\alpha} = 0$). This is characteristic of the singlets being isolated from the rest of the system. 
The $S^z$–$S^z$ correlation matrix (Fig.\,\ref{fig:le}(e)) further confirms the absence of any correlations outside individual unit cells in the system. The correlations seen along the diagonal of the correlation matrix are for the parallel spin-$1$ sites oriented antiferromagnetically to the other parallel site within the same unit cell.

Next, in the phase $P7$, the system forms a perfect spin-1 chain ( Fig. \ref{fig:le}(j)); however, this phase is not a Haldane phase, despite exhibiting a finite string-order parameter, as shown in Fig. \ref{fig:de-en-ee}(c), because the static properties of the system are inconsistent with a Haldane phase.
To start, we note that this is a product state similar to $P3$; however, instead of singlets on the dimers, it is composed of one of the triplet states, $\frac{1}{\sqrt{2}}\left(\ket{\uparrow 0}+\ket{0\uparrow}\right)$.
The phase transition from $P3$ to $P7$ occurs due to a level crossing that changes the ground state of the system. At higher magnetic fields, the energy of the local triplet state $\frac{1}{\sqrt{2}}\left(\ket{\uparrow 0}+\ket{0\uparrow}\right)$ on each orthogonal dimer is lowered relative to that of the singlet state $\frac{1}{\sqrt{3}}\left(\ket{\uparrow\downarrow}+\ket{\downarrow\uparrow}-\ket{00}\right)$.

The characteristics of this product state of triplets are reflected in local observables Fig.\,\ref{fig:le}(b) as well as in the spatial distribution of entanglement entropy in Fig.\,\ref{fig:le}(d). The SQN gives a fully spin-$1$ system, but the magnetization do not go to zero for both the parallel sites and the orthogonal dimers in the bulk. For the parallel sites, the magnetization saturates to a value of $0.5$. The entanglement entropy of the system is also $0$ across the orthogonal dimer bonds, and the parallel bonds saturates to slightly below $0.75$.
Furthermore, the $S^z$–$S^z$ correlation matrix depicted in Fig.\,\ref{fig:le}(f) indicates the presence of ferromagnetic correlations throughout the whole system, in contrast to an exponentially decaying correlations found in Haldane phases. Therefore, $P7$ is not a Haldane phase, and the finite string order is not indicative of a topological phase.

Due to the product-state nature of the phases $P3$ and $P7$, and the zero entanglement at the bonds connecting the orthogonal dimers, an excitation created on the orthogonal dimer bonds cannot propagate to its nearest neighbor. 
Thus, the DSF for both of these phases $P3$ and $P7$ are expected to be similar and dispersionless.
For demonstration, we calculate DSF for the phase $P7$ using Equation \ref{eqn:DSF_t},
\begin{equation}
\begin{split}
    S^{t}(k,\omega) & = \sum\limits_{n=-\infty}^{\infty} e^{-ikn} \int_{-\infty}^{\infty} dt\;e^{i\omega t} \\
    & \;\langle\psi|\;S^{+}_n(t)S_0^{-}(t)\;
    + \;S^{-}_n(t)S_0^{+}(t)\;|\psi\rangle
    \label{eqn:DSF_t}
\end{split}
\end{equation}
with the index $0$ on an orthogonal dimer, and plotted the results in Fig.\,\ref{fig:le}(g).
As the excitations are localized, we see a very flat band, which has a field-dependent gap of $\omega/J\approx 0.5$ at $h/J=1.53$, due to the energy difference between the singlet and triplet state in the system.


To calculate the dynamics of the low lying excitations on parallel sites using the DSF Equation \ref{eqn:DSF}, one of the parallel sites can be chosen (we chose the left parallel site), which results
in a dispersionless branch centered at the point $k=\pi$. However, two bands on the left and right can be seen, at around the point $k \approx \frac{3\pi}{4}$ and extending to the left and $k \approx \frac{5\pi}{4}$ extending to the right respectively. The excitation on the parallel site can still delocalize to the neighboring parallel site, but is isolated from other unit cells by the presence of the orthogonal dimers. We expect the difference in Figures \ref{fig:le} (g) and (h) to be attributed to this fact.

\subsection{Effective Haldane Phases}\label{ehp}

\captionsetup{justification=justified, font=small}
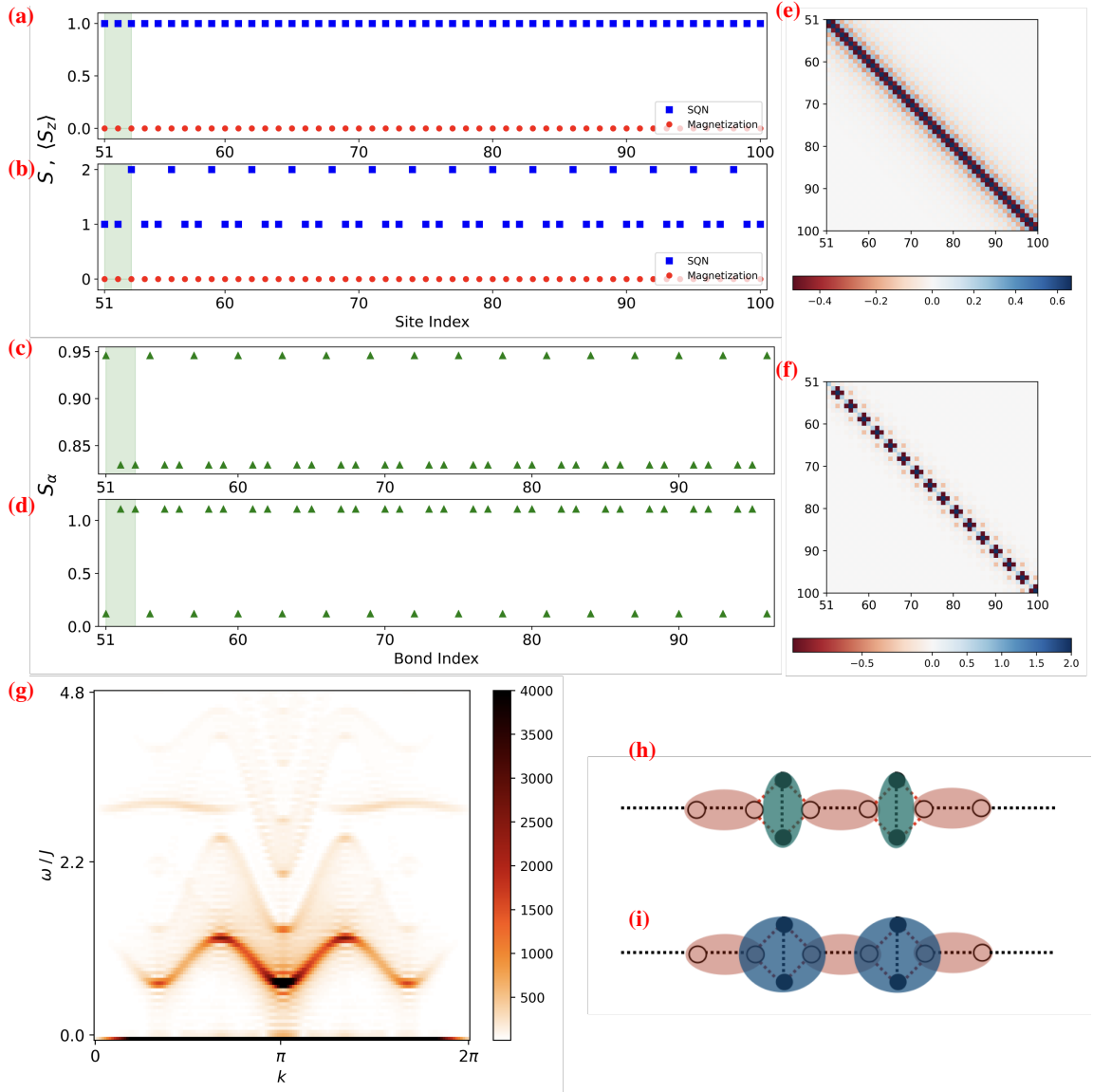
\begin{figure*}
    \centering
    \begin{overpic}[width=0.9\textwidth,grid=false]{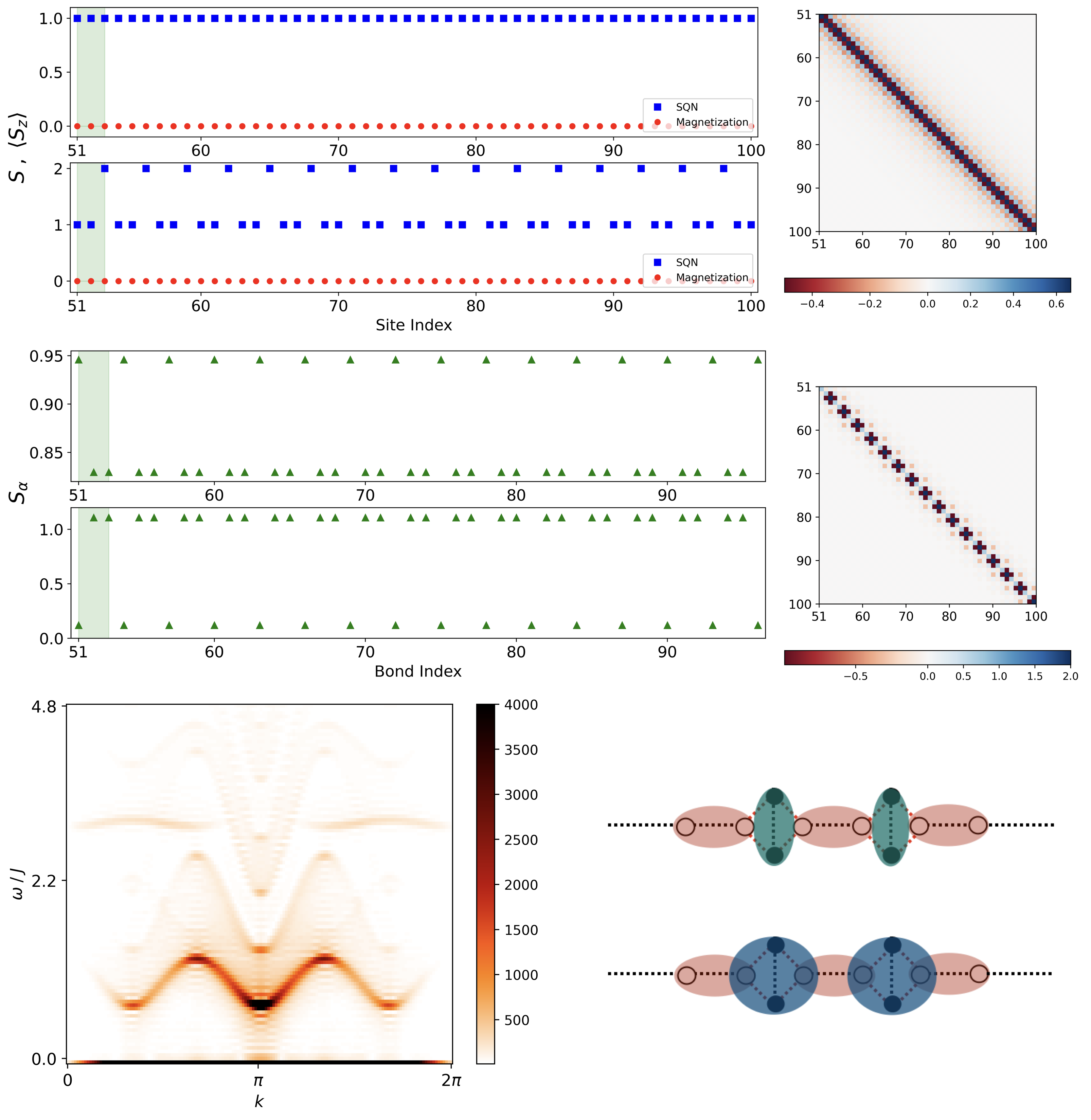} 
    \put(-2,98.5){\textcolor{red}{\textbf{(a)}}} 
    \put(-2,84.5){\textcolor{red}{\textbf{(b)}}}
    \put(-2,68){\textcolor{red}{\textbf{(c)}}}
    \put(-2,53.5){\textcolor{red}{\textbf{(d)}}}

    \put(68.5,99){\textcolor{red}{\textbf{(e)}}}
    \put(68.5,66){\textcolor{red}{\textbf{(f)}}}
    
    \put(-2,36.5){\textcolor{red}{\textbf{(g)}}}

    \put(55,31){\textcolor{red}{\textbf{(h)}}}
    \put(55,15.5){\textcolor{red}{\textbf{(i)}}}

  \end{overpic}
    \caption{a),b) The SQN and magnetization calculations for the phase $P4$ and $P5$ respectively. Similar to Fig. \ref{fig:le}, the figure begins with a parallel site. c),d) The entanglement entropy for calculations for the phase $P4$ and $P5$ respectively. The first bond, at index $51$ is the bond connecting parallel sites. e),f) The correlation matrices for the phase $P4$ and $P5$ respectively. g) The DSF calculation for the phase $P5$ using Equation \ref{eqn:DSF} (with $\alpha=\beta=z$) on a parallel site. h,i) The representative diagram for the phases $P4$ and $P5$ respectively. The green ovals are spin-$1$ state of orthogonal dimers and the blue ovals represent a quadrumer state with the orthogonal dimer in spin-$2$ state.}
    \label{fig:eh}
\end{figure*}



Due to exponentially decaying correlations and the absence of local magnetic ordering, the topologically non-trivial Haldane phase cannot be detected through experiments that probe local order parameters.
While the Haldane phase was originally proposed as the ground state of an integer-spin chain described by the isotropic Heisenberg model with antiferromagnetic interactions, it can also emerge in quasi-one-dimensional spin systems within specific interaction regimes.
An effective Haldane phase for the system in this study appears in the $P4$ regime (Fig.\,\ref{fig:de-en-ee}(a)), where the SQN of the orthogonal dimer bond reduces to $1$, effectively mapping the quasi-one-dimensional chain into a one-dimensional spin-$1$ chain (see Fig.\,\ref{fig:eh}(a)).
Moreover, Fig.\,\ref{fig:eh}(a) shows that the bulk magnetization vanishes, indicating the absence of local magnetic order, while Fig.\,\ref{fig:eh}(c) reveals a nearly uniform spatial entanglement entropy distribution—both characteristic signatures of the Haldane phase.
Exponential decay in correlation in Fig.\,\ref{fig:eh}(e) further supports this identification.
This phase also exhibits a finite string order parameter (see Fig.\,\ref{fig:de-en-ee}(c)), a hallmark of the Haldane phase's topological nature.
The schematic of the spin chain in this effective Haldane phase is shown in Fig.\,\ref{fig:eh}(h).

Remarkably, we find that the phase $P5$ exhibits characteristics of the Haldane phase.
While for sufficiently large frustration strength $J'/J$, $P5$ is expected to evolve into a plaquette-ordered (or quadrumer) phase, it surprisingly retains a finite string order parameter—indicative of underlying Haldane-like topological order.
The magnetization and SQN values (Fig.,\ref{fig:eh}(b)) indicate that the spin chain in this phase behaves as an integer-spin chain with alternating spin values, and exhibits no local magnetic order, like the Haldane phase.
Although the entanglement across the dimer bonds is very low and the correlations decay exponentially, both remain finite (Fig.\,\ref{fig:eh}(d), (f)).
This is further supported by the presence of non-localized excitations, as evidenced by the non-dispersive nature of the DSF\,\ref{fig:eh}(g).
Interestingly, the band structure resembles that of the Haldane phase, particularly exhibiting a gap at the $k=\pi$ point. 
However, the overall structure is more complex due to the presence of three additional bands above the lowest one.
Notably, there is a flat band at energy $\omega/J=3.06$. 
This complex band structure arises from the coexistence of two phases within $P5$, the quadrumer phase and the Haldane phase.
The schematic of the system in phase $P5$ is represented in Fig.\,\ref{fig:eh}(i).

\captionsetup{justification=justified, font=small}
\begin{figure*}
    \centering
    \begin{overpic}[width=0.9\textwidth,grid=false]{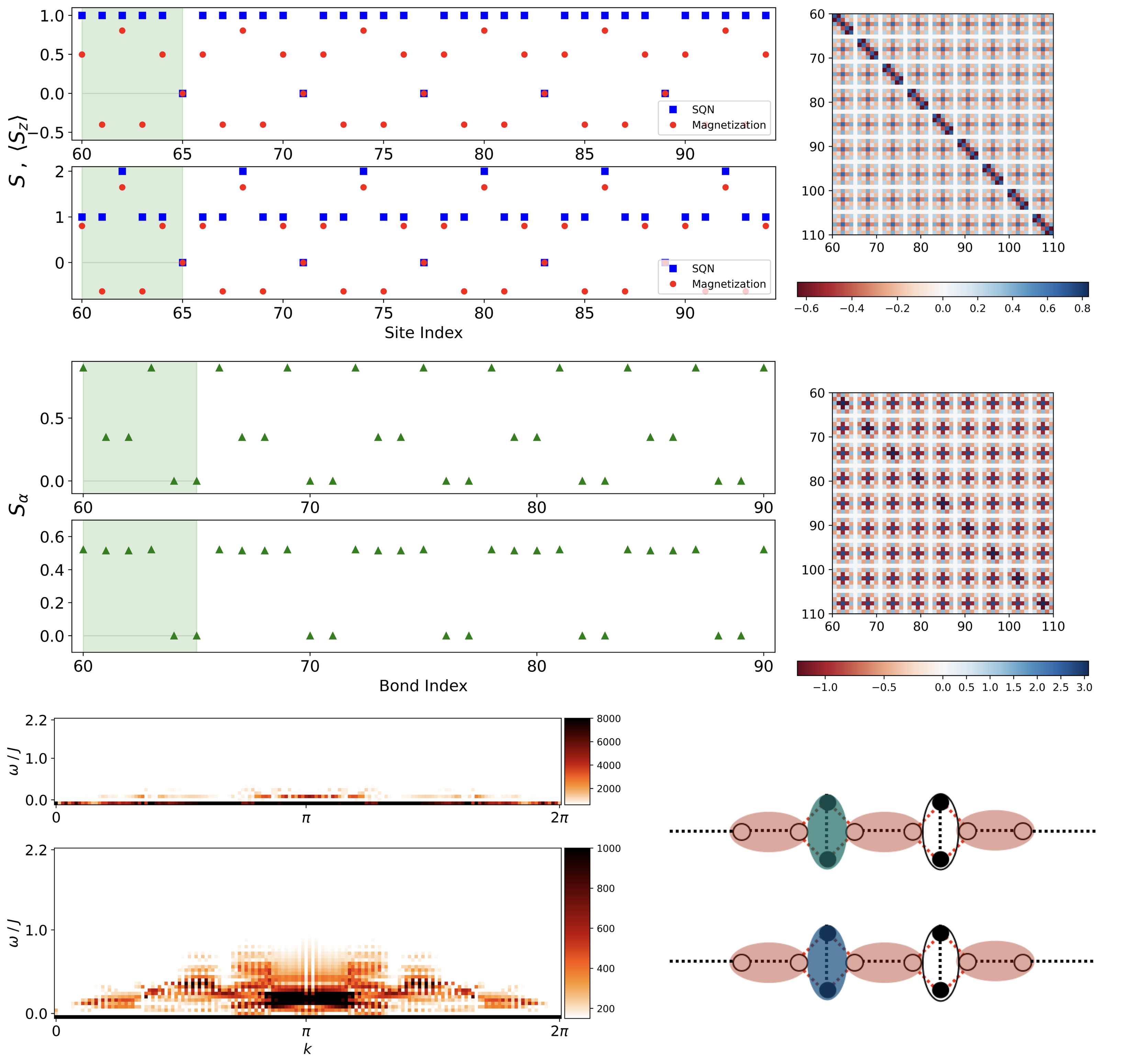}
    \put(-2,92.5){\textcolor{red}{\textbf{(a)}}}
    \put(-2,77.5){\textcolor{red}{\textbf{(b)}}}

    \put(-2,61){\textcolor{red}{\textbf{(c)}}}
    \put(-2,47){\textcolor{red}{\textbf{(d)}}}

    \put(69,93.5){\textcolor{red}{\textbf{(e)}}}
    \put(69,60.5){\textcolor{red}{\textbf{(f)}}}

    \put(-2,29.5){\textcolor{red}{\textbf{(g)}}}
    \put(-2,18){\textcolor{red}{\textbf{(h)}}}

    \put(63,25){\textcolor{red}{\textbf{(i)}}}
    \put(63,13){\textcolor{red}{\textbf{(j)}}}
  \end{overpic}
    \caption{a,b) The SQN and magnetization calculation of the phase $CP8$ and $CP9$, starting and ending with parallel sites, taken inside the bulk. c,d) The entanglement entropy of the bulk, with the first bond index is for the bond connecting two parallel sites, for $CP8$ and $CP9$ respectively. e,f) The correlation profiles for $CP8$ and $CP9$. g,h) The DSF calculation done using Equation \ref{eqn:DSF} (with $\alpha=\beta=z$) on a parallel site for the phases $CP8$ and $CP9$ respectively. i,j) The representative diagrams for the phases $CP8$ and $CP9$ respectively. The green (blue) ovals denote spin-$1$ (spin-$2$) states of orthogonal dimers. Hollow oval denote orthogonal dimer singlets.}
    \label{fig:ch}
\end{figure*}

\subsection{Clustered Phases and Clustered Haldane Phase}\label{chp}


Systems in clustered phases are essentially composed of product states of small entangled spin clusters, which remain disentangled from the rest of the system due to the formation of singlets on the dimers at the end of the spin clusters.
We identify three distinct cluster phases $CP8$, $CP9$, and $CP15$ as indicated in the phase diagram Fig.\,\ref{fig:de-en-ee}(a).
Notably, the phases $CP8$ and $CP9$ are of particular interest due to their proximity to the effective Haldane phase $P4$ in the phase diagram, suggesting that they may inherit certain characteristic features of the Haldane phase.
A cluster phase exhibiting characteristics of the Haldane phase is referred to as the clustered-Haldane phase, which was first introduced by Katarina {\it et al.}\cite{Katarina}. 
Further analysis indicates that the cluster phase $CP8$ corresponds to a clustered-Haldane phase, while $CP9$ represents a trivial clustered phase.

We begin by discussing the phase $CP8$.
The SQN, magnetization, and correlation profiles of the $CP8$ phase are shown in Fig.\,\ref{fig:ch}(a), (c), and (e), respectively.
Notably, the SQN values within a cluster (indicated by the green region) effectively correspond to a spin-$1$ chain, with a magnetization of $0.5$ at the boundary parallel sites—key characteristics of the spin-$1$ Haldane chain. 
Furthermore, the finite connected correlation observed within the cluster can be attributed to the distinction between intra-cluster and inter-cluster correlations, as shown in Fig.\,\ref{fig:ch}(e).
Moreover, the spatial distribution of entanglement shown in Fig.\,\ref{fig:ch}(c) confirms the presence of clustering, as the entanglement vanishes at the cluster boundaries.
Within each cluster, the entanglement is relatively uniform, and the degree of non-uniformity observed in Fig.\,\ref{fig:ch}(c) closely resembles that of the effective Haldane phase depicted in Fig.\,\ref{fig:eh}(c).
The flat DSF shown in Fig.\,\ref{fig:ch}(g) suggests that excitations are localized and cannot propagate between clusters.

However, no signatures indicative of the Haldane phase are observed in the $CP9$ phase. 
Specifically, the SQN values within the cluster do not correspond to a spin-$1$ chain, nor is the magnetization of the parallel sites near the at the cluster boundary fractional, as shown in Fig.\,\ref{fig:ch}(b).
Additionally, the connected correlation within the cluster vanishes, which can be understood as the intra-cluster and inter-cluster correlation remains similar, as indicated in Fig.\,\ref{fig:ch}(f). 
Interestingly, in Fig.\,\ref{fig:ch}(d), the entanglement distribution within the cluster is remarkably uniform; however, this alone is not a significant signature of the Haldane phase.
Furthermore, the flat DSF for $CP8$ is indicative of a cluster-like phase.
However, we observe a broadening and slight dispersion in the DSF of the $CP9$ phase, which can be attributed to two possible reasons: first, the flat band is highly degenerate, and this degeneracy is partially lifted; second, some low-lying excitations break the singlet formations at the ends of the clusters, enabling weak inter-cluster hopping and resulting in weak dispersion.


\section{\label{sec:level1}Summary and Conclusion}

We have studied the spin-$1$ orthogonal dimer model using the Density Matrix Renormalization Group (DMRG) method.
To ensure an unbiased DMRG calculation, we implemented a straightforward basis transformation that incorporates the symmetries of the system.
By analyzing the first- and second-order derivatives of the ground state energy, we mapped out a rich phase diagram of the spin-$1$ orthogonal dimer chain and estimated the phase boundaries (see Fig.~\ref{fig:de-en-ee}(a)).

It is worth noting that machine learning-based approaches for automatic and more precise identification of phase boundaries have been proposed in recent studies \cite{automaticlearning}, which could be promising for future work.
Among the various identified phases, we focus on low-entanglement phases ($P3$, $P7$), effective Haldane phases ($P4$, $P5$), clustered phases ($CP9$), and clustered Haldane phases ($CP8$).
To characterize and distinguish these phases, we compute several diagnostics, including spin quantum numbers, entanglement entropy, correlation functions, string order parameters, and the dynamical spin structure factor.

The low-entanglement phase $P3$ is a simple product state composed of singlets, specifically $\frac{1}{\sqrt{3}}\left(\ket{\uparrow\downarrow}+\ket{\downarrow\uparrow}-\ket{00}\right)$, localized on the dimer bonds.
As the magnetic field increases, $P3$ undergoes a transition into another product state formed by triplets, given by $\frac{1}{\sqrt{2}}\left(\ket{\uparrow 0}+\ket{0\uparrow}\right)$.
Upon further increase in the field, this intermediate phase eventually evolves into the fully polarized state $P6$.
Due to their product-state nature, these phases exhibit low entanglement entropy and feature a nearly flat band structure, reflecting the strong localization of excitations.

Furthermore, we demonstrate that phase $P4$ is an effective Haldane phase, where the system effectively maps onto a chain of spin-$1$ degrees of freedom.
Interestingly, phase $P5$—which corresponds to a quadrumerized regime in the limit $J'/J \gg 1$—also displays clear signatures of the Haldane phase.
In particular, the dynamical structure factor (DSF) in phase $P5$ closely resembles that of the Haldane phase, featuring a characteristic low-energy excitation band, along with a higher-energy flat band indicative of the underlying quadrumer structure.
This suggests that both the characteristics of Haldane and quadrumer phase coexist in phase $P5$.

We also identify several clustered phases ($CP8$, $CP9$, $CP15$), which are low-entanglement phases composed of product states localized on small spin clusters.
Among them, $CP8$ and $CP9$ are of particular interest due to their proximity to the effective Haldane phase $P4$.
A recent study by Katarina et al.~\cite{Katarina} demonstrated that certain cluster phases can exhibit features reminiscent of the Haldane phase, referring to them as clustered Haldane phases.
In our analysis, we find that $CP8$ qualifies as a clustered Haldane phase: the spin clusters in this phase behave as effective spin-$1$ chain and exhibit a characteristic fractionalized edge excitation with spin-$\frac{1}{2}$—a hallmark of the Haldane phase.

In summary, our study reveals that the phase diagram of the spin-$1$ orthogonal dimer chain is remarkably rich, featuring a variety of quantum phases.
In particular, we identify several topologically non-trivial phases that are connected to the Haldane phase.

\section{Acknowledgements}

It is a pleasure to thank Andreas Honecker for useful discussions. This work was partially supported by funding from the Ministry of Education, Singapore, through a Tier 1 grant RG 138/22. We would like to acknowledge the High Performance Computing Centre of Nanyang Technological Centre, Singapore, for providing the computing resources, facilities, and services that have contributed significantly to this work.


\appendix

\section{Phase Diagram Construction}\label{pd-appendix}

\begin{figure*}
    \centering
    \includegraphics[width=1\linewidth]{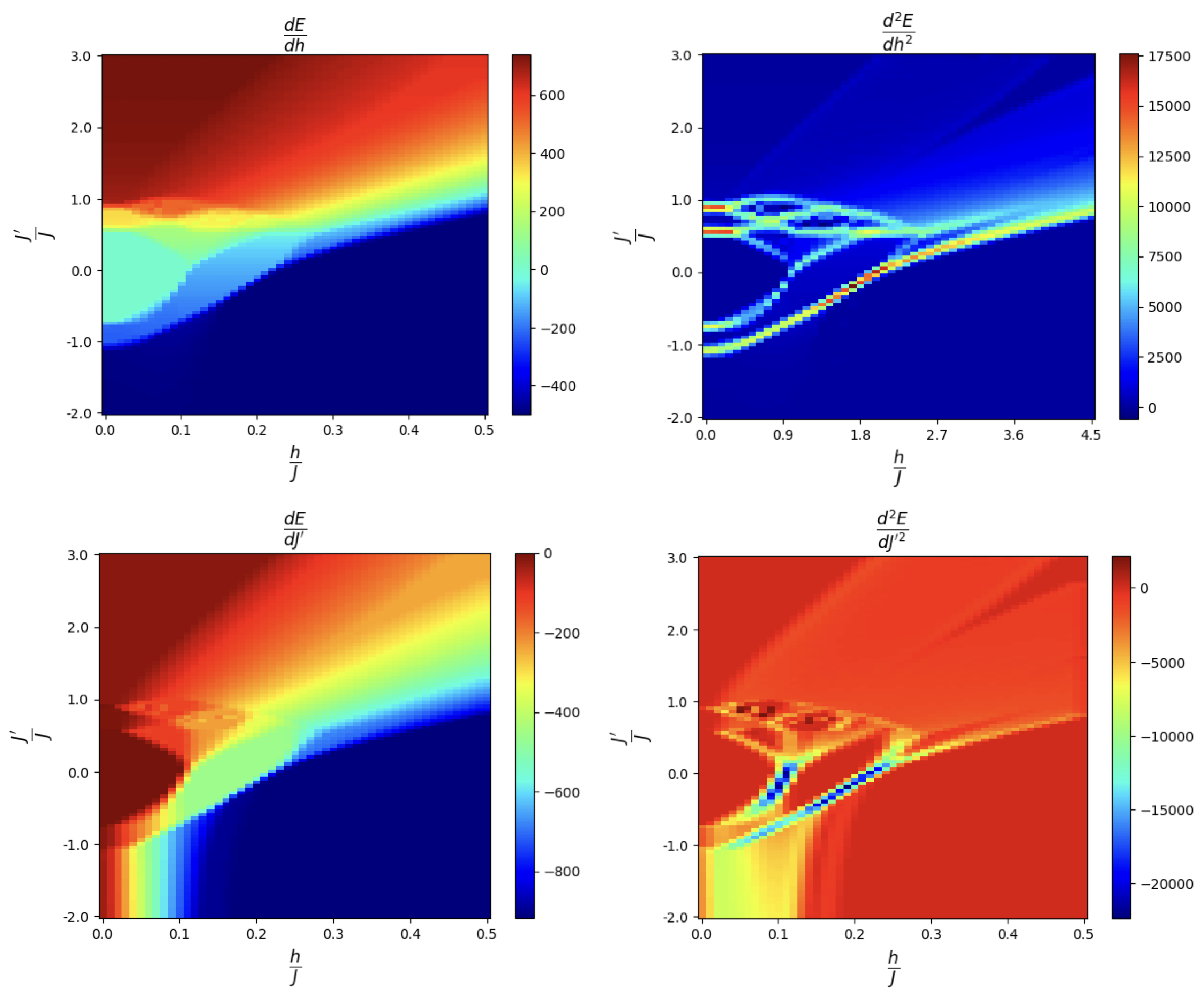}
    \caption{The first (top left) and second order (top right) energy derivatives with respect to the magnetic field. The first (bottom left) and second order (bottom right) energy derivatives with respect to the interaction J'}
    \label{fig:edboth}
\end{figure*}

To construct the phase diagram, we have calculated the first and second order derivatives of the energy with respect to the interaction $J'$ and the magnetic field $h$, shown in Figure \ref{fig:edboth}

After obtaining the derivatives of the energy, the outlines of each phases were traced to produce the phase diagram in the main text. The peaks shown in the second derivatives are indicative of second order transitions, which are outlined as black lines on the phase diagram in the main text. 

The lines where a first order transition would happen is also apparent where the gradient of the energy changes, without any peaks in the second order derivative.

\section{On Phases with Similar SQN Structure}\label{diff}

\vspace{3em}
\begin{figure*}
    \centering
    \begin{overpic}[width=0.75\textwidth,grid=false]{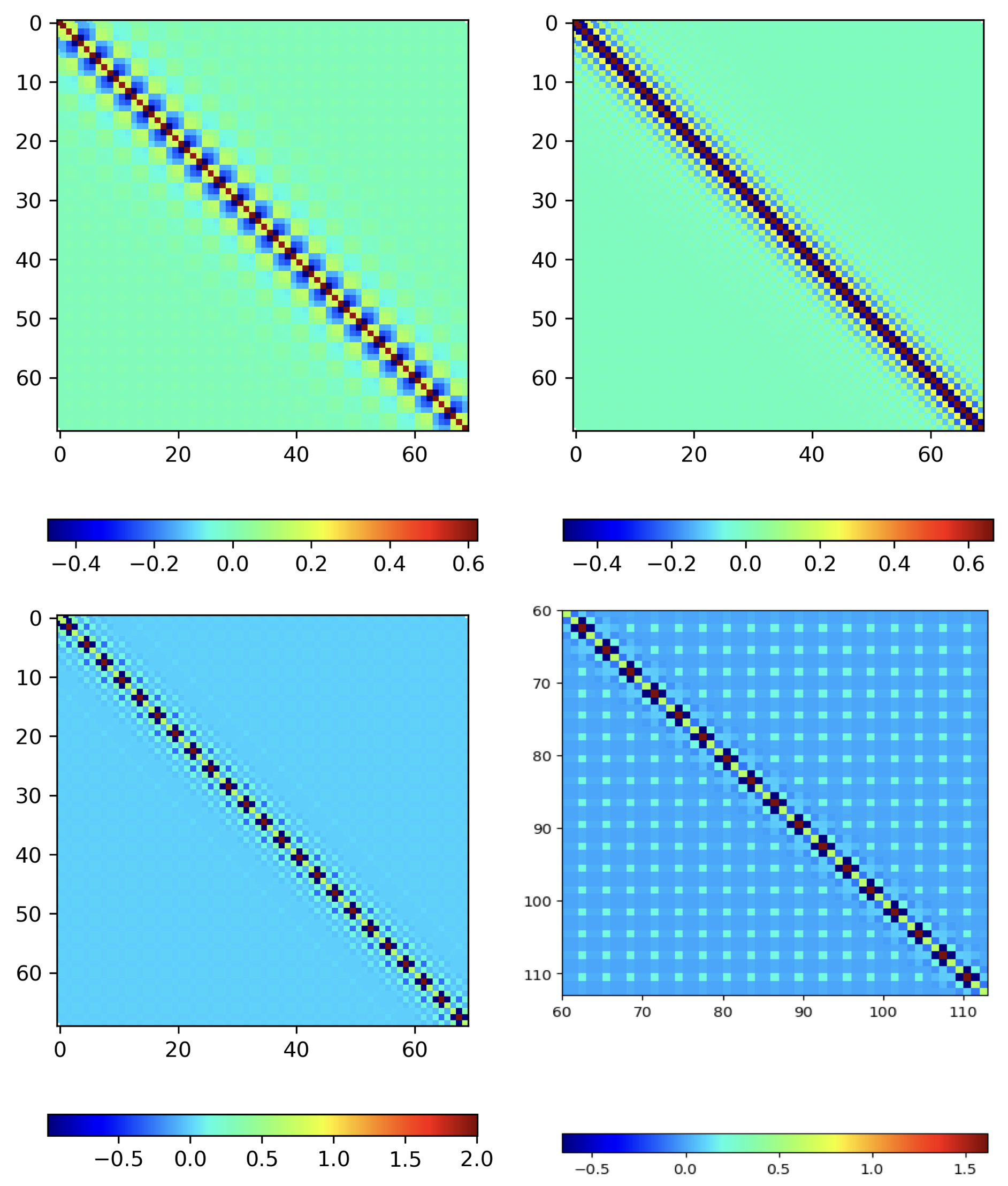} 
    \put(41,97){\textcolor{red}{\textbf{(a)}}} 

    \put(41,46){\textcolor{red}{\textbf{(b)}}}
    \end{overpic}
    \caption{Comparison of phases with similar SQN profiles in Figure \ref{fig:de-en-ee}$a$. Shown are for figures differentiated by different signs of $J'$ ( a) $P2$ against $P4$) , and differentiated via a first order quantum phase transition ( b) $P5$ against $P12$)}
    \label{fig:compare}
\end{figure*}

From the phase diagram in Figure \ref{fig:de-en-ee}$a$, some of the phases have the same SQN profile in each unit cell, for example for the phases $P1$ and $P6$, but the difference in the phases are apparent when looking at other static properties. Phases with similar SQN are mostly found when they are neighbouring each other via a first order transition, for example the phases $P1,P6$ and $P5,P12,P13,P14$, but the tuning of the interaction from antiferromagnetic to ferromagnetic $J'$ also shows similar phases such as the phase $P2,P4$ and $P1,P5$.

Comparing the static properties of $P5$ and $P12$ (bottom two figures, Figure \ref{fig:compare}$b$), we see the correlation profiles of both the phases are different. A similar investigation on the phases with similar SQN profiles but separated by second order transitions instead yields the same outcome (phases $P2$ and $P4$ in the top two figures in Figure \ref{fig:compare}$a$). The 15 numbered phases shown in the phase diagram are ultimately different, and can be differentiated with the inclusion of more static properties than just the SQN.

\clearpage



\providecommand{\noopsort}[1]{}\providecommand{\singleletter}[1]{#1}%

\end{document}